%% file: paper.tex
\pdfoutput=1
\documentclass[pdflatex,sn-mathphys-num]{sn-jnl}% Math and Physical Sciences Numbered Reference Style
\usepackage{graphicx}%
\usepackage{multirow}%
\usepackage{amsmath,amssymb,amsfonts}%
\usepackage{amsthm}%
\usepackage{mathrsfs}%
\usepackage[title]{appendix}%
\usepackage{xcolor}%
\usepackage{textcomp}%
\usepackage{manyfoot}%
\usepackage{booktabs}%
\usepackage{algorithm}%
\usepackage{algorithmicx}%
\usepackage{algpseudocode}%
\usepackage{listings}%
\usepackage{cleveref}
\usepackage{tcolorbox}
\usepackage{listings}
\usepackage{xcolor}
\usepackage{etoolbox}

\newlength{\bibitemsep}
\newlength{\bibparskip}
\let\oldthebibliography\thebibliography
\renewcommand\thebibliography[1]{%
  \oldthebibliography{#1}%
  \setlength{\parskip}{\bibitemsep}%
  \setlength{\itemsep}{\bibparskip}%
}
\makeatletter
\patchcmd{\@maketitle}{\removelastskip\vskip24pt}{\removelastskip\vskip10pt}{}{}
\patchcmd{\@maketitle}{\removelastskip\vskip24pt}{\removelastskip\vskip10pt}{}{}
\patchcmd{\@maketitle}{\removelastskip\vskip36pt}{\removelastskip\vskip12pt}{}{}
\makeatother

\tcbuselibrary{breakable,skins}
\newtcolorbox{promptbox}[1][]{
    enhanced,
    breakable,
    colback=gray!5,
    colframe=gray!50,
    fonttitle=\bfseries\ttfamily\small,
    title={#1},
    boxrule=0.5pt,
    arc=2pt,
    left=8pt,
    right=8pt,
    top=6pt,
    bottom=6pt,
    fontupper=\ttfamily\small,
}

\begin{document}

\title[SHIELDS: Automating OS Hardening with Iterative Multi-Agent Remediation]{SHIELDS: Automating OS Hardening with Iterative Multi-Agent Remediation}

% L3Harris authors
\author[1]{\fnm{Andrew} \sur{Hamara}}\email{hamaraa@acm.org}
\author*[1]{\fnm{Dwight} \sur{Horne}}\email{dwighthorne1@acm.org}
\author[1]{\fnm{Aldehir} \sur{Rojas}}\email{aldehirr@acm.org}

% A&M authors
\author[2]{\fnm{Timothy} \sur{Kurniawan} \textsuperscript{\ddag}}\email{timothykurniawan16@tamu.edu}
\author[2]{\fnm{Sophie} \sur{Lamothe} \textsuperscript{\ddag}}\email{smlamothe@tamu.edu}
\author[2]{\fnm{Vishal} \sur{Suresh} \textsuperscript{\ddag}}\email{suresh06192004@tamu.edu}
\author[2]{\fnm{Nicholas} \sur{Turoci} \textsuperscript{\ddag}}\email{n2rowc@tamu.edu}
\author[2]{\fnm{Lawrence} \sur{Wong} \textsuperscript{\ddag}}\email{lawrencewong@tamu.edu}

% affiliations
\affil[1]{\orgname{L3Harris Technologies}}
\affil[2]{\orgname{Texas A\&M University}}

% footnote for alphabetical order for A&M students
\begingroup
\renewcommand\thefootnote{}\footnotetext{\textsuperscript{\ddag}These authors are listed alphabetically.}
\addtocounter{footnote}{-1}
\endgroup

% footnote for ITAR
\begingroup
\renewcommand\thefootnote{}\footnotetext{This document consists of information that is not defined as controlled technical data under ITAR Part 120.33 or technology under EAR Part 772.}
\addtocounter{footnote}{-1}
\endgroup

\abstract{Security misconfigurations remain a leading cause of OS-level compromise, and manually keeping systems compliant with standards like Defense Information Systems Agency (DISA) Security Technical Implementation Guides (STIGs) is a tedious and expensive process. Existing compliance automation tools can reduce some of this burden, but they depend on static, pre-written corrective actions. In this paper, we introduce SHIELDS, a multi-agent system that uses large language models (LLMs) to approach OS hardening as an iterative, feedback-driven process. Instead of applying fixed remediations, SHIELDS continuously proposes fixes and refines them based on feedback from target system execution and validation scans. We evaluate the system across multiple virtual machine configurations using six contemporary LLMs ranging from 20B to 400B parameters, and find that SHIELDS successfully remediates up to 73\% of scan findings. Our results also suggest that success in this setting depends less on model size (parameter count) than on effective tool use and information gathering, paving a practical path toward reducing the burden of security compliance in environments where compute is limited or security and privacy needs drive local model use.}


\maketitle

%%%%%%%%%%%%%%%%%%%%%%%%%%%%%%%%%%%%%%%%%%%%%%%%%%%%%%%%%%%%%%%%%%%%%%%%%%%%%%%%%%%%%%%%%

\section{Introduction}\label{sec:introduction}

Operating system (OS) misconfigurations remain one of the most consequential sources of cybersecurity risk due to their prevalence and potential impact. IBM's 2024 Cost of a Data Breach Report found that the average breach reached a record cost of \$4.88 million~\cite{ibm2024breach}. Furthermore, the Verizon 2024 Data Breach Index Report revealed that 68\% of breaches involved non-malicious human mistakes such as misconfigurations, and that vulnerability exploitation as an initial attack vector nearly tripled to 14\% of all breaches~\cite{verizon2024dbir}. By 2025, that figure climbed further to 20\%~\cite{verizon2025dbir}.

To reduce the risk posed by configuration weaknesses, frameworks such as Defense Information Systems Agency (DISA) Security Technical Implementation Guides (STIGs) provide standardized OS hardening baselines. However, since DISA publishes updates on a quarterly cycle and each update potentially invalidates prior remediation work, manual scanning and remediation are required several times per year. SteelCloud estimates that manual STIG compliance requires approximately 16 hours per server each year, and organizations may require as many as 20 full-time engineers dedicated exclusively to compliance maintenance~\cite{steelcloud2023case}. At this scale, manual system compliance is untenable.

Existing automation tools partially address this burden. Projects like ComplianceAsCode~\cite{complianceascode}, Ansible Lockdown~\cite{ansiblelockdown}, and PowerSTIG~\cite{powerstig} provide pre-authored remediations for certain issues, while OpenSCAP evaluates systems against checklists and can generate fix scripts for known findings~\cite{redhat_openscap}. These tools work well for rules with existing fixes, but every STIG rule requires a handcrafted remediation that may vary by OS version and must be updated whenever DISA publishes new guidance. When a remediation attempt fails, OpenSCAP can report the failure, but it does not provide mechanisms for adaptive retries or context-aware reasoning about \emph{why} a proposed fix did not work or alternative remediation strategies. In other words, current tools lack the ability to adapt when a remediation fails or when no predefined fix exists, potentially leaving many findings unresolved. Recent advances in large language models (LLMs) suggest a promising path toward addressing these limitations. Unlike static tools, LLMs can iteratively reason about system state and adapt based on feedback. Although prior work has demonstrated these capabilities in system administration tasks and vulnerability repair~\cite{genkubesec,vrpilot,llmpatch,managing}, their effectiveness for achieving compliance with real-world STIG profiles remains largely unexplored.

To address this gap, we present \emph{Security Hardening with Intelligent Expert Language Driven Systems (SHIELDS)}, a system that treats STIG compliance as a feedback-driven remediation loop. Rather than relying on predefined fixes, SHIELDS iteratively generates, executes, and refines remediations based on live scan results. The system is implemented as a multi-agent architecture that separates triage, remediation, validation, and safety enforcement. Critically, the Remediation Agent incorporates feedback from OS command execution and scanner feedback into subsequent remediation attempts, enabling adaptive compliance workflows that extend beyond the static remediation capabilities of existing tools.

Some key contributions of this paper include:
\begin{enumerate}
    \item We present SHIELDS, a multi-agent architecture and system available under an MIT license for autonomous STIG remediation that combines scanning, reasoning, remediation, safety review, and validation.
    \item We introduce a feedback-driven remediation workflow that iteratively adapts corrective actions based on OS command execution results and scanner validation rather than relying solely on static, predefined fixes.
    \item We evaluate SHIELDS with benchmark data across multiple Linux configurations using six primarily open-weights LLMs from 20B to 400B parameter sizes and analyze the relationships between model characteristics and remediation performance.
\end{enumerate}

%%%%%%%%%%%%%%%%%%%%%%%%%%%%%%%%%%%%%%%%%%%%%%%%%%%%%%%%%%%%%%%%%%%%%%%%%%%%%%%%%%%%%%%%%

\section{Related Work}\label{relatedwork}

Existing approaches to operating system hardening can be broadly divided into two categories: rule-based compliance automation and AI-enabled security reasoning systems. The first category includes established compliance tools such as OpenSCAP~\cite{redhat_openscap}, ComplianceAsCode~\cite{complianceascode}, Ansible Lockdown~\cite{ansiblelockdown}, and PowerSTIG~\cite{powerstig}. These frameworks can automate portions of the compliance process by mapping individual STIG requirements to predefined remediation scripts. While this can be highly effective for known configurations, in essence they operate as expert systems whose behavior is constrained by manually developed rule sets. Consequently, coverage is limited to previously authored remediations, and adapting to new STIG revisions requires continual human maintenance. Recent work by Liu et al.~\cite{liu2024automatic} extends automated repair beyond compliance scripts by proposing a localize-fix-validate workflow for network misconfigurations. However, that approach relies on predefined change operators and historical repair patterns for network configuration environments rather than direct OS hardening against STIG findings.

The second category encompasses recent applications of large language models (LLMs) and autonomous agents to cybersecurity tasks. Prior studies have demonstrated that LLMs can assist with vulnerability triage, software repair, penetration testing support, infrastructure management, and operational troubleshooting~\cite{genkubesec,vrpilot,llmpatch,managing}. For example, Wang et al.~\cite{Wang2025LLM} studied an LLM-supported collaborative vulnerability remediation process among security technicians, users, and LLMs, and showed that LLM assistance can sometimes reduce remediation time, while also imposing practical limitations when proposed solutions are too generic or difficult to validate. These systems leverage reasoning and decision support to perform tasks that are difficult to encode through static rules, but they stop short of a compliance-focused autonomous remediation loop to address standardized hardening requirements. Toward this end, multi-agent architectures have emerged as a promising paradigm for decomposing complex workflows into specialized planning, execution, and validation roles~\cite{talebirad2023multi,agenticmindset}.

A recent survey by Rokade and Bhakulkar~\cite{AIcyber2026} examined the state of AI-driven STIG automation within DevSecOps environments. Their analysis concluded that traditional automation achieves inconsistent compliance coverage and highlighted semantic policy interpretation, contextual reasoning, and adaptive remediation as key research challenges. The authors argued that AI-native compliance systems could potentially overcome limitations of traditional methods but noted the absence of empirical validation demonstrating such capabilities in operational environments. SHIELDS occupies the intersection of these two research directions. Unlike conventional compliance frameworks, SHIELDS does not rely exclusively on predefined remediation libraries. In contrast to prior LLM security systems, its objective is not vulnerability analysis alone or software repair, but direct compliance remediation against standardized hardening requirements. By combining agentic AI reasoning with scanner-based validation, SHIELDS implements an autonomous scan-analyze-fix-verify loop capable of adapting remediation strategies based on observed outcomes, an empirical validation of promising opportunities recently identified in the literature.

%%%%%%%%%%%%%%%%%%%%%%%%%%%%%%%%%%%%%%%%%%%%%%%%%%%%%%%%%%%%%%%%%%%%%%%%%%%%%%%%%%%%%%%%%

\section{Methodology}\label{sec:methodology}

This section provides an overview of our methods, focusing on our proposed multi-agent system and experimental benchmarking setup.

\subsection{Multi-Agent Architecture}\label{sec:multiagent}

Rather than relying on a single agent to handle processing, our pipeline splits the work across four specialized agents. We separate these responsibilities across specialized agents to reduce prompt complexity, improve modularity, and reduce the likelihood of errors from hallucinations or cascading hallucinations in accordance with evolving best practices in agentic AI system design~\cite{agenticmindset}. Additionally, separation of responsibilities enabled the introduction of independent review stages prior to application of changes to the target system as a component of overall risk management. Here, we describe the purpose of each agent with respect to the system as a whole. All agent prompts are shown in~\cref{appendix_prompts} for transparency.

\paragraph{Triage Agent.} The Triage Agent acts as a preprocessing filter by deciding what to do with each finding: remediate it, discard it, or flag it for human review. A hard-coded filter runs first and catches partition and file system rules, which are always classified as too dangerous for automation. This small number of findings is excluded from automation because incorrect modifications may render the system unbootable or result in data loss.

\paragraph{Remedy Agent.} The Remedy Agent is the central component of our system. It has access to three tools: command execution (\texttt{run\_cmd}), reading of files (\texttt{read\_file}), and file modification (\texttt{write\_file}), all of which are available once the Remedy Agent proposes a fix. These tools allow the agent to inspect the system state, modify configuration files, and apply remediation actions directly on the target host. The Remedy Agent is allowed up to three remediation attempts per finding by default, a value selected based on preliminary experiments that showed diminishing returns beyond three iterations.

\paragraph{Review Agent.} The Review Agent evaluates fix quality of the proposed remediation given by the Remedy Agent in a single chat completion. It produces a security impact score (1--10), an optimality assessment, and lists any concerns. Critically, if the Review Agent rejects the remediation plan for any reason, control is returned to the Remedy Agent with updated context.

\paragraph{QA Agent.} The QA Agent is also a single-completion evaluator, though it focuses on system-wide safety rather than individual fix quality. It analyzes proposed system changes and considers potential operational impacts such as service disruption, dependency changes, and unintended configuration drift, then issues a recommendation: \emph{Approve}, \emph{Rollback}, or \emph{Investigate}. Similar to the Review Agent, if the QA Agent deems the proposed fix as too risky, control is returned to the Remedy Agent with context.

\subsection{System Overview}\label{sec:system}

With the individual agents defined, we now discuss their broader goal in our proposed SHIELDS system. The pipeline, shown in \cref{fig:system}, operates as follows:

% results figure
\begin{figure}[h]
\centering
\makebox[\textwidth][c]{\includegraphics[width=1.2\textwidth]{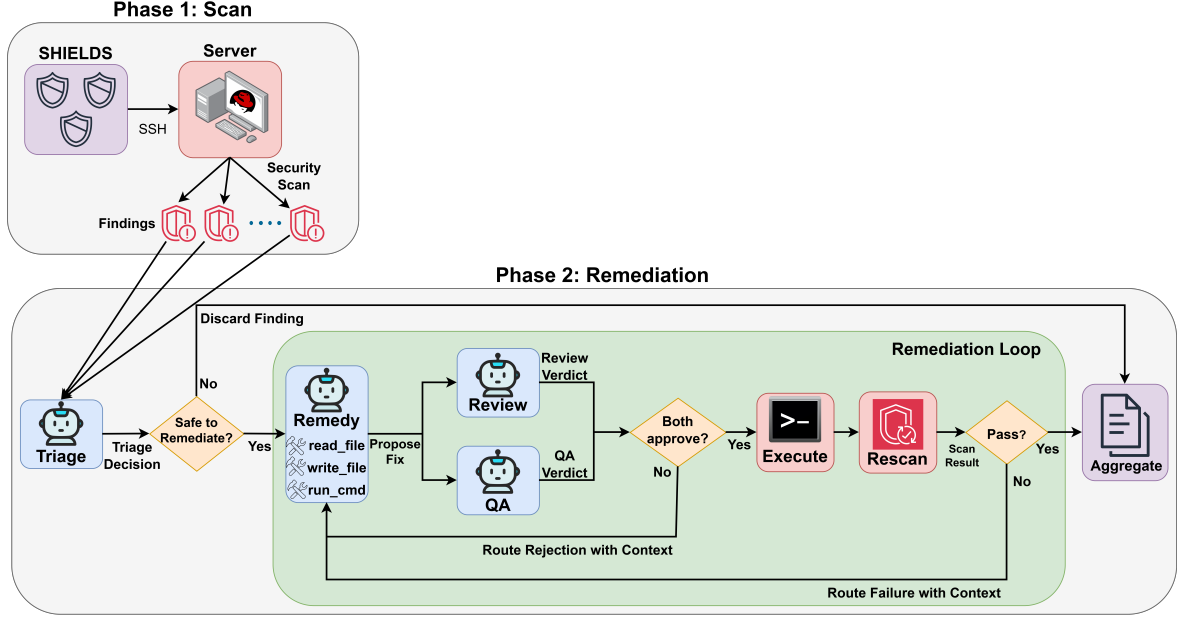}}
\caption{The full SHIELDS remediation pipeline.} \label{fig:system}
\end{figure}

\begin{enumerate}

    \item \textbf{Scan.} SHIELDS starts in Phase 1 by remotely connecting to a RHEL/Rocky machine \emph{(Server in \cref{fig:system})} and initiating a system-wide \emph{Security Scan} on the target machine. This results in a set of \emph{Findings}, each of which is passed to the \emph{Triage Agent} block.

    \item \textbf{Triage.} The Triage Agent makes a decision for the finding \emph{(Triage Decision)}. If the Triage Agent decides that the finding is not safe to remediate, the finding is discarded and sent to \emph{Aggregate}, meaning that we mark it as unresolved and log this result for metrics. If the finding is safe to remediate, control is given to the \emph{Remedy Agent}, marking the start of the \emph{Remediation Loop}, highlighted in green in \cref{fig:system}.

    \item \textbf{Remedy loop} (up to three attempts). If the Triage Agent concludes that the finding is safe to remediate, the Remedy Agent proposes a fix \emph{(Propose Fix)}. The \emph{Review} and \emph{QA} Agents each evaluate the fix independently, and together their verdicts create a dual approval gate \emph{(Both approve?)}. If both approve, the Remedy Agent implements its plan by executing commands \emph{(Execute)}. After this, a single-finding OpenSCAP scan is run for the finding in question \emph{(Rescan)}. If the Review Agent or QA Agent rejects, or if the single finding scan fails, the rejection feedback is appended to the context for the prompt with the next remediation attempt. If the scan passes, the finding gets aggregated as a success \emph{(Aggregate)}.

    \item \textbf{Aggregation.} Once all findings are either discarded, remediated, or attempted three times with failure, statistics are computed across all findings and an Ansible playbook is generated from successful fixes.
\end{enumerate}

\subsection{Models Under Evaluation}\label{sec:models}

To evaluate this workflow, we benchmark six models spanning 20B to 400B parameter range. Table~\ref{tab:remedy_models} lists the models used in our evaluation. We focus primarily on open-weight models to enable reproducibility and transparency. We additionally include the closed-weight Inception Mercury 2 model~\cite{mercury} since benchmarks suggest that it is competitive with models in this size class while requiring lower inference cost. Although testing utilized OpenRouter.ai for convenience due to lack of hardware needed to run an instance of all AI models simultaneously, the system leverages standard APIs and can also use local model providers.

\subsection{Experimental Environment}\label{sec:testimages}

For reproducibility, three baseline (golden) test images were established as VMs for the benchmark testing and every experimental iteration began with a clean VM snapshot. Three Rocky Linux 9 virtual machine images were constructed to represent small, medium, and large remediation workloads, containing approximately 25, 80, and 250 initial OpenSCAP findings respectively. All experiments were performed against identical STIG profiles and OpenSCAP scan configurations. Experiments were executed for each AI model and VM size combination to produce the complete suite of benchmark results. Additionally, the SHIELDS project source code is available on GitHub under an MIT license for further testing and flexible extensibility~\cite{GitHubSHIELDS}.

Throughout all experiments, the same agent prompts, workflow, tool access, and remediation policies were used across AI models. Consequently, differences in performance primarily reflect differences in model behavior rather than differences in the underlying remediation pipeline. However, note also that differences in overall success can be due to differences in performance of an individual agentic role (e.g., the Remedy Agent proposing superior or lower probability fixes), or due to the unique collaborative emergent behavioral interactions of the entire agentic team powered by a given AI model (e.g., a highly accurate Remedy Agent combined with an overly cautious QA Agent). Therefore, observed performance differences are primarily attributable to model-specific behavior within the context of the multi-agent SHIELDS architecture.

% models table
\begin{table}[h]
\centering
\caption{Models evaluated in our benchmarks along with their total parameter count, active parameter count during inference, and context window.}
\label{tab:remedy_models}
\begin{tabular}{l|ccc}
\hline
\textbf{Model} & \textbf{Parameters} & \textbf{Active Parameters} & \textbf{Context Window} \\
\hline
{Trinity Large 400B~\cite{trinity}}    & 398B  & 13B  & 128k \\
{Google Gemma 4 26B A4B~\cite{gemma}}     & 25.2B & 3.8B & 256k \\
{NVIDIA Nemotron 3 Super~\cite{nemotron}} & 120B  & 12B  & 256k \\
{OpenAI gpt-oss-20b~\cite{gptoss}}        & 21B   & 3.6B & 128k \\
{OpenAI gpt-oss-120b~\cite{gptoss}}      & 117B  & 5.1B & 128k \\
{Inception Mercury 2~\cite{mercury}}      & 117B  & 5.1B & 128k \\
\hline
\end{tabular}
\end{table}

%%%%%%%%%%%%%%%%%%%%%%%%%%%%%%%%%%%%%%%%%%%%%%%%%%%%%%%%%%%%%%%%%%%%%%%%%%%%%%%%%%%%%%%%%

\section{Results}\label{results}

% results figure
\begin{figure}[h]
\centering
\makebox[\textwidth][c]{\includegraphics[width=1.2\textwidth]{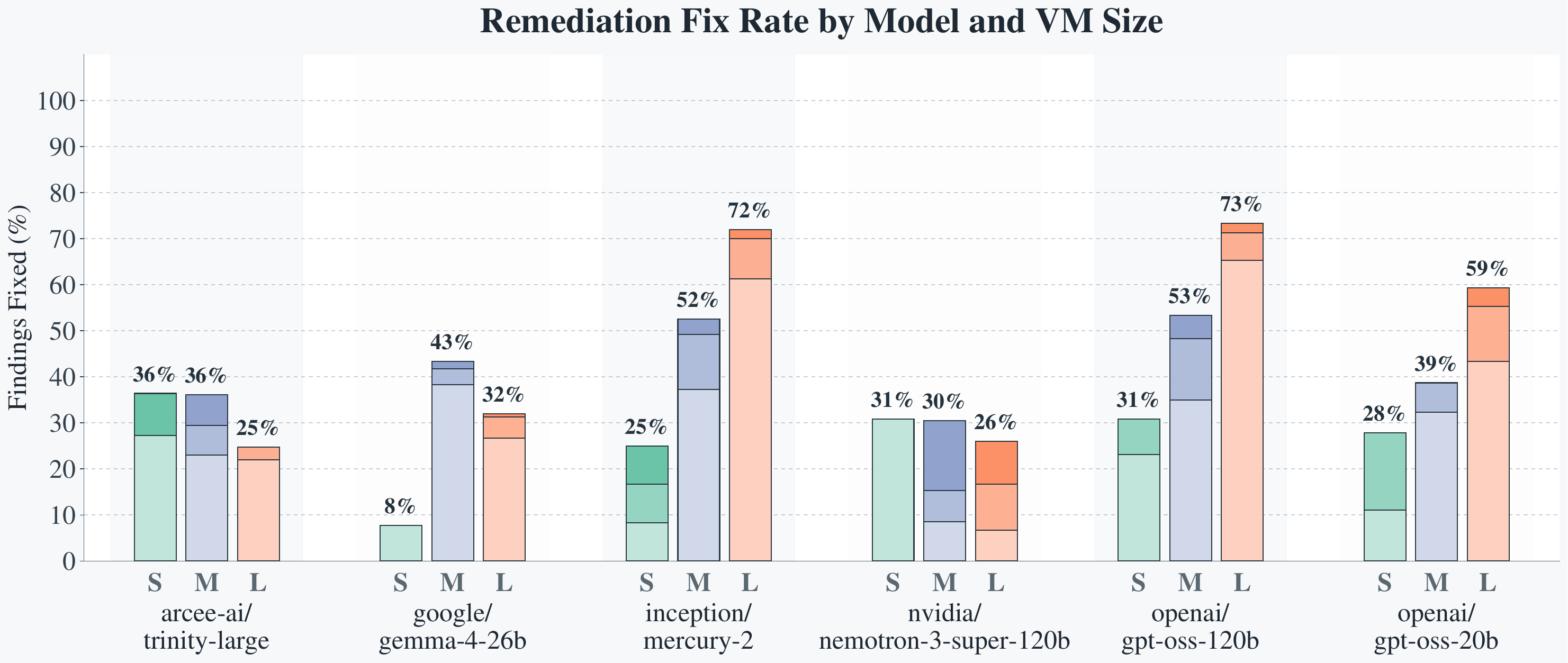}}
\caption{Remediation fix rate for all models used in our evaluation. Results for each model are reported on three VMs: small (green), medium (blue), and large (orange). Within each bar, hues increase in darkness for each pass through all findings.}\label{fig:results_fig}
\end{figure}

Benchmark results for each model are presented visually in~\cref{fig:results_fig} and in tabular form in~\cref{tab:results}.

Across all benchmark configurations, remediation rates varied substantially by model, with gpt-oss-120b and Mercury 2 consistently achieving the highest percentages of resolved findings for the medium and high workload test cases. Interestingly, we find that model size (parameter count) does not necessarily imply high performance in this task. For example, Trinity Large --- the largest model used in our evaluations by a large margin --- was inferior to gpt-oss-20b for two of three test cases, a model almost 20x smaller. To that point, the two highest-performing models --- Mercury 2 and gpt-oss-120b --- excel at roughly 120b parameters, yet Nemotron 3 exhibited weaker success rates at the same size. These results suggest that dataset selection or post-training may be more indicative of accuracy in this task than scaling alone. Since this task requires the Remedy Agent to gather lots of system-wide information, models that are more adept at tool calling may have an advantage because of their superior ability to gather context, and based on the detailed results we hypothesize that this tool-use proficiency contributed more substantially to strong performance than pre-trained knowledge. 

We also observe that most remediations typically occur on the first pass through findings, and that in some cases, \emph{no} findings are remediated on passes two and three. There are, however, some cases where roughly equal amounts of findings are remediated on all three passes (e.g., Mercury 2 on the small VM). We experimented with adding two additional passes during early experiments, but saw diminishing returns at passes four and five, both of which typically had no additional fixes. Because of this, we conclude that using three passes in production would be a reasonable trade-off between speed and remediation percentage.

Finally, with Mercury 2 and both gpt-oss variants, we observe a positive correlation between VM size and percentage of findings fixed. To explain this, we suggest that all the VMs likely share similar core weaknesses, while the larger VMs also pick up easier, more mechanical findings (e.g, file permissions or package upgrades) that these models can easily resolve.

%%%%%%%%%%%%%%%%%%%%%%%%%%%%%%%%%%%%%%%%%%%%%%%%%%%%%%%%%%%%%%%%%%%%%%%%%%%%%%%%%%%%%%%%%

\section{Future Work}\label{sec:future}

Given our finding that model size does not necessarily imply accuracy for system-wide OS configuration, future work will include fine-tuning smaller models and utilization of models specialized for tool calling. Since our runs were relatively time-consuming due to long reasoning during the planning phase, integration with lightweight tool-calling models such as functiongemma~\cite{functiongemma} to offload some work is a promising path toward making SHIELDS more practical for real-world deployment.

We also note that models of this size are often highly prompt-sensitive. Prompting highly-specialized agents such as our Triage and QA agents is particularly difficult, and failure to properly describe their intended behavior can easily lead to them being overly strict or lenient. Future work could include hand-picked examples that can be used as a test dataset for prompt engineering to enable more rigorous refinement.

Given that performance may depend on interactions between agents as much as the quality of proposed fixes from the Remedy Agent, future work should further examine the emergent behaviors of AI models and impact of assigning specialized models to individual agent roles. For example, a light-weight tool-calling model may be sufficient for remediation planning, while larger reasoning models may be better suited for safety review and validation. This design may also have implications for cost characteristics and practicality for offline use cases. Additionally, practical extensions such as expansion of the human-in-the-loop approval feature, a formalized rollback system, and Review/QA Agents that consider the unique software intended to execute on the target system are natural roadmap priorities.

\section{Conclusion}\label{sec:conclusion}

We have presented SHIELDS, a multi-agent AI system that approaches OS cybersecurity hardening and compliance as an iterative, feedback-driven process rather than a static remediation problem. By combining specialized agents for triage, remediation, validation, and safety review with security scanner guided feedback, SHIELDS is capable of autonomously generating and refining corrective actions for real-world DISA STIG findings. Across multiple Rocky Linux virtual machine configurations and six contemporary LLMs, SHIELDS successfully remediated up to 73\% of identified scan findings when equipped with AI models in the 20B to 120B range. Our results further suggest that remediation effectiveness is driven less by model scale than by a model's ability to gather information and use tools effectively, indicating that practical compliance automation may be achievable without relying on frontier-scale models.

More broadly, this work demonstrates that autonomous security hardening can extend beyond traditional rule-based compliance frameworks. By treating compliance as a closed-loop scan-analyze-fix-verify process, SHIELDS provides an empirical example of how agentic AI systems can adapt remediation strategies in response to real system feedback. We believe these results represent an important step toward reducing the operational burden of security compliance and enabling more intelligent, adaptive approaches to establishing and maintaining hardened systems.

\section{Acknowledgements}\label{sec:acknowledgements}

The student research team thanks L3Harris Technologies (Company) for sponsoring and mentoring this capstone project at Texas A\&M University. This report of project results should not be interpreted as Company endorsement of, or affiliation with, the SHIELDS software or Texas A\&M University. Any views in the paper or project contents, expressed or implied, comprise personal views of the authors. Company makes no representations or warranties of any kind, express or implied, related to the paper content or SHIELDS software.

\bibliography{sn-bibliography}% common bib file
%% if required, the content of .bbl file can be included here once bbl is generated
%%\input sn-article.bbl

%%%%%%%%%%%%%%%%%%%%%%%%%%%%%%%%%%%%%%%%%%%%%%%%%%%%%%%%%%%%%%%%%%%%%%%%%%%%%%%%%%%%%%%%%

\backmatter

\begin{appendices}

%%%%%%%%%%\newpage
\section{Agent Prompts}\label{appendix_prompts}

This section contains the prompts we use for our Remedy, Review, QA, and Triage agents in all experiments.

%%%%%%%%%%%%%%%%%%%%%%%%%%%%%%%%%%%%%%%%%%%%%%%%%%%%%%%%%%%%%%%%%%%%%%%%%%%%%%%%%%%%%%%%%

\subsection{Remedy Agent}
\begin{promptbox}[Remedy Agent System Prompt]
You are an adaptive remediation agent on Rocky Linux / RHEL.
Follow this workflow STRICTLY:
1. PLAN: In your first response, describe your proposed fix
(which files, which values, which commands).
2. REVIEW: Call review\_plan with your full plan description
BEFORE executing any commands.
3. APPLY: Only after review\_plan returns approved=true, use
run\_cmd / read\_file / write\_file to apply the fix.
4. If review\_plan returns approved=false, read the feedback,
revise your approach, and call review\_plan again. You have a
maximum of 3 review attempts - if all are rejected, proceed
with your best plan using the execution tools.
EXECUTION RULES:
\begin{itemize}
\item{One command at a time. Do NOT chain with \&\& or ;}
\item{Always use read\_file before modifying a config file.}
\item{This is Rocky Linux/RHEL - use dnf (not apt), systemctl (not service).}
\item{Do NOT duplicate config lines. Modify existing values in-place with sed.}
\item{Do NOT modify sshd\_config or firewall rules in ways that block SSH.}
\item{stderr may contain SSH banners - check exit\_code and stdout instead.}
\item{A verification scan runs automatically after your session ends.}
\end{itemize}
\end{promptbox}

%%%%%%%%%%%%%%%%%%%%%%%%%%%%%%%%%%%%%%%%%%%%%%%%%%%%%%%%%%%%%%%%%%%%%%%%%%%%%%%%%%%%%%%%%

%%%%%\newpage
\subsection{Review Agent}
\begin{promptbox}[Review Agent System Prompt]
You are a security remediation reviewer. Evaluate the following proposed remediation plan for quality and correctness.

\textbf{Vulnerability}
\begin{itemize}
\item ID: \{vulnerability.id\}
\item Title: \{vulnerability.title\}
\item Severity: \{vulnerability.severity\}
\item Description: \{vulnerability.description or "(none)"\}
\item Recommendation: \{vulnerability.recommendation or "(none)"\}
\end{itemize}

\textbf{Triage}
\begin{itemize}
\item Risk level: \{triage.risk\_level\}
\item Reason: \{triage.reason\}
\end{itemize}

\textbf{Proposed Remediation Plan}

(NOTE: This plan has NOT been executed yet. Evaluate whether it WOULD resolve the vulnerability if executed correctly.)

\{remediation\_attempt.llm\_verdict.message\}

\textbf{Previous Review History (if applicable)}

\begin{itemize}
\item Review \#\{review\_index\}: approve=\{previous\_verdict.approve\}, score=\{previous\_verdict.security\_score\}
\item Concerns raised: \{previous\_verdict.concerns (up to 5, semicolon-separated)\}
\item Improvements requested: \{previous\_verdict.suggested\_improvements (up to 5, semicolon-separated)\}
\item Feedback: \{previous\_verdict.feedback (truncated to 200 characters)\}
\end{itemize}

Check whether the current fix addresses the issues raised in previous reviews.

Respond with a single JSON object (no markdown, no extra text) with these exact keys:

finding\_id (string), is\_optimal (bool), approve (bool), feedback (string or null), \\
concerns (list of strings), suggested\_improvements (list of strings), \\
security\_score (integer 1--10 or null), best\_practices\_followed (bool).

\textbf{IMPORTANT:}

Set approve=true if the proposed plan would functionally resolve the vulnerability when executed.  
Only set approve=false if the plan is actively harmful, introduces new security risks, or would fail to address the vulnerability.  
Do NOT reject because commands have not been run yet — this is a plan review before execution.

Use is\_optimal, concerns, and suggested\_improvements to note areas for improvement without blocking the fix from proceeding.
\end{promptbox}

%%%%%%%%%%%%%%%%%%%%%%%%%%%%%%%%%%%%%%%%%%%%%%%%%%%%%%%%%%%%%%%%%%%%%%%%%%%%%%%%%%%%%%%%%

%%%%%\newpage
\subsection{QA Agent}
\begin{promptbox}[QA Agent System Prompt]
You are a pragmatic QA validation agent for Linux security remediation. Your role is to verify that a remediation did NOT introduce serious harm to the system.  You have access to 1 tool: 1. run\_cmd: Run system health checks (systemctl status, service checks, log analysis) 
Your validation checklist:
\begin{itemize}
\item{Verify critical services are still running (sshd, auditd, firewalld, etc.)}
\item{Confirm system is still accessible (SSH works)}
\item{Detect any major side effects or unintended changes}
\end{itemize}
SAFETY GUIDELINES:
\begin{itemize}
\item{Mark safe=true if the system is still functional and critical services are running.}
\item{Minor warnings in logs, non-critical service restarts, or cosmetic issues should NOT cause a failure.}
\item{Mark safe=false ONLY if critical services are down, the system is unreachable, or the remediation clearly broke something important.}
\item{Security remediations are expected to change configurations - that alone is not a side effect.}
\item{When done, call verdict with safe=true if the system is functional and healthy - safe=false only if serious issues are detected}
\item{Include detailed message explaining findings}
\end{itemize}
\end{promptbox}

%%%%%%%%%%%%%%%%%%%%%%%%%%%%%%%%%%%%%%%%%%%%%%%%%%%%%%%%%%%%%%%%%%%%%%%%%%%%%%%%%%%%%%%%%

%%%%%\newpage
\subsection{Triage Agent}
\begin{promptbox}[Triage Agent System Prompt]
Classify this OpenSCAP finding into exactly one category:

\texttt{safe\_to\_remediate}, \texttt{requires\_human\_review}, \texttt{too\_dangerous\_to\_remediate}.

\textbf{Output Format}

Return ONLY JSON for this schema:

\begin{lstlisting}[basicstyle=\ttfamily\small, breaklines=true]
{
  "finding_id": string,
  "rule_id": string,
  "category": "safe_to_remediate" | "requires_human_review" | "too_dangerous_to_remediate",
  "confidence": number between 0 and 1,
  "rationale": string,
  "risk_factors": [string, ...],
  "safe_next_steps": [string, ...],
  "requires_reboot": boolean,
  "touches_authn_authz": boolean,
  "touches_networking": boolean,
  "touches_filesystems": boolean,
  "estimated_complexity": "low" | "medium" | "high"
}
\end{lstlisting}

\textbf{Finding}

\begin{itemize}
\item finding\_id: \{vulnerability.id\}
\item rule\_id: \{vulnerability.title\}
\item severity: \{vulnerability.severity\}
\item title: \{vulnerability.title\}
\item description: \{vulnerability.description (truncated to 900 chars)\}
\item recommendation: \{vulnerability.recommendation (truncated to 900 chars)\}
\end{itemize}

\textbf{Complexity Estimation}

\begin{itemize}
\item low: simple configuration edit, package install, or service toggle (1-5 minutes)
\item medium: multi-file changes, service restart, moderate validation (5-15 minutes)
\item high: filesystem-wide scans, FIPS mode, kernel parameters, large permission sweeps, or extensive validation (15+ minutes)
\end{itemize}
\end{promptbox}

%%%%%%%%%%%%%%%%%%%%%%%%%%%%%%%%%%%%%%%%%%%%%%%%%%%%%%%%%%%%%%%%%%%%%%%%%%%%%%%%%%%%%%%%%

%\newpage
\hspace{2cm}
\section{Additional Tables and Figures}\label{appendix:morestuff}
\subsection{Benchmark Results Table}\label{appendix:results}
\include{remediation_passes_table}

\end{appendices}

\end{document}

%% file: remediation_passes_table.tex
% Requires \usepackage{booktabs}
\begin{table}
\centering
\small
\setlength{\tabcolsep}{6pt}
\renewcommand{\arraystretch}{1.15}
\caption{Detailed remediation results by model and VM configuration.}
\label{tab:results}
\begin{tabular}{llrrrr}
\toprule
Model & VM Config & Findings & Pass 1 & Pass 2 & Pass 3 \\
\midrule
arcee-ai/trinity-large-preview & Small & 11 & 3 (27.3\%) & 3 (27.3\%) & 4 (36.4\%) \\
 & Medium & 61 & 14 (23.0\%) & 18 (29.5\%) & 22 (36.1\%) \\
 & Large & 150 & 33 (22.0\%) & 37 (24.7\%) & 37 (24.7\%) \\
\midrule
google/gemma-4-26b-a4b-it & Small & 13 & 1 (7.7\%) & 1 (7.7\%) & 1 (7.7\%) \\
 & Medium & 60 & 23 (38.3\%) & 25 (41.7\%) & 26 (43.3\%) \\
 & Large & 150 & 40 (26.7\%) & 47 (31.3\%) & 48 (32.0\%) \\
\midrule
inception/mercury-2 & Small & 12 & 1 (8.3\%) & 2 (16.7\%) & 3 (25.0\%) \\
 & Medium & 59 & 22 (37.3\%) & 29 (49.2\%) & 31 (52.5\%) \\
 & Large & 150 & 92 (61.3\%) & 105 (70.0\%) & 108 (72.0\%) \\
\midrule
nvidia/nemotron-3-super-120b & Small & 13 & 4 (30.8\%) & 4 (30.8\%) & 4 (30.8\%) \\
 & Medium & 59 & 5 (8.5\%) & 9 (15.3\%) & 18 (30.5\%) \\
 & Large & 150 & 10 (6.7\%) & 25 (16.7\%) & 39 (26.0\%) \\
\midrule
openai/gpt-oss-120b & Small & 13 & 3 (23.1\%) & 4 (30.8\%) & 4 (30.8\%) \\
 & Medium & 60 & 21 (35.0\%) & 29 (48.3\%) & 32 (53.3\%) \\
 & Large & 150 & 98 (65.3\%) & 107 (71.3\%) & 110 (73.3\%) \\
\midrule
openai/gpt-oss-20b & Small & 18 & 2 (11.1\%) & 5 (27.8\%) & 5 (27.8\%) \\
 & Medium & 62 & 20 (32.3\%) & 24 (38.7\%) & 24 (38.7\%) \\
 & Large & 150 & 65 (43.3\%) & 83 (55.3\%) & 89 (59.3\%) \\
\bottomrule
\end{tabular}
\end{table}